\documentclass[english,aps,prc,0superscriptaddress,twocolumn,longbibliography,notitlepage]{revtex4-1}
\usepackage{amsmath,amssymb,multirow,bm}
\usepackage{soul}
\usepackage{epsfig}
\usepackage{graphicx}
\usepackage{amsmath}
\usepackage{lipsum}
\usepackage{color}
\usepackage{float}
\makeatother

\usepackage{babel}
\usepackage[colorlinks,linkcolor=blue,anchorcolor=blue,citecolor=blue,urlcolor=blue,filecolor=black]{hyperref}

\newcommand{\beq}{\begin{equation}}
\newcommand{\eeq}{\end{equation}}
\newcommand{\bea}{\begin{eqnarray}}
\newcommand{\eea}{\end{eqnarray}}
\usepackage{graphicx}
\preprint{}

\begin{document}

\title{A Hierarchical Bayesian Analysis of Neutron-Skin Thicknesses and Implications for the Symmetry-Energy Slope}
\author{Adrian Azizi$^1$, Carlos A. Bertulani$^{1,2}$, and Carlos Davila$^1$} 
\affiliation{$^1$Department of Physics, East Texas A\&M University, Commerce, Texas 75429, USA}
\affiliation{$^2$Technische Universit\"at Darmstadt,  Institut f\"ur Kernphysik, 64289, Darmstadt, Germany}
\email{carlos.bertulani@etamu.edu}

\date{\today}

\begin{abstract}
Neutron-skin thicknesses provide a sensitive probe of the isovector sector of the nuclear equation of state and its density dependence, commonly characterized by the symmetry-energy slope parameter $L$. A wide variety of experimental and observational methods have been used to extract neutron skins, ranging from hadronic and electromagnetic probes of finite nuclei to inferences from neutron-star observations. Each approach carries distinct theoretical and systematic uncertainties, complicating global interpretations and obscuring genuine physical trends.
In this work we present a hierarchical Bayesian framework for the statistically consistent synthesis of heterogeneous neutron-skin constraints. The neutron-skin thickness is modeled as a smooth latent function of isospin asymmetry and nuclear size, while method-dependent bias parameters and intrinsic nuisance widths are introduced to account for unmodeled experimental and theoretical systematics. 
Focusing on the tin isotopes, we infer probabilistic neutron-skin trends from $^{100}$Sn to $^{140}$Sn, finding minimal uncertainties near stability and increasing uncertainties toward the proton-rich and neutron-rich extremes. We assess the consistency of nuclear energy-density functionals and obtain conditional constraints on the symmetry-energy parameters. The resulting posterior exhibits a pronounced compression of the symmetry-energy slope parameter $L$, reflecting the dominant sensitivity of neutron skins to sub-saturation symmetry pressure.
We demonstrate that our hierarchical Bayesian framework provides robust and transparent constraints on the sub-saturation isovector sector of the nuclear equation of state.\end{abstract}

\maketitle

\section{Introduction}
The neutron-skin thickness of a nucleus,
\begin{equation}
\Delta r_{np} = \sqrt{\langle r_n^2 \rangle} - \sqrt{\langle r_p^2 \rangle},
\end{equation}
quantifies the spatial separation between neutron and proton density distributions and provides a sensitive probe of the isovector sector of the nuclear equation of state (EoS). The formation of a neutron skin reflects a competition between surface tension, which favors compact neutron-proton distributions, the nuclear symmetry energy, which penalizes isospin asymmetry in the bulk, and Coulomb repulsion, which reshapes proton densities and indirectly modifies the isovector density profile. From a bulk perspective, neutron skins can be interpreted as manifestations of the symmetry pressure that drives excess neutrons toward the nuclear surface in neutron-rich systems.

A substantial body of theoretical work has established a strong correlation between neutron-skin thicknesses in medium-heavy and heavy nuclei and the density dependence of the symmetry energy, particularly its slope parameter $L$ near nuclear saturation density $\rho_0$
 \cite{Brown2000,TypelBrown2001,Furnstahl2002,Centelles2009,RocaMaza2011}. Since the pressure of neutron-rich matter is governed by the density derivative of the energy per particle, neutron skins are far more sensitive to $L$ than to the symmetry energy at saturation, $J$. As a result, neutron-skin measurements provide a quantitative bridge between finite-nucleus observables and astrophysical properties of neutron stars, including their radii and tidal deformabilities \cite{HorowitzPiekarewicz2001,LattimerPrakash2007,Horowitz2014,Annala2018}.
 
 Over the past two decades, neutron-skin information has been extracted using a wide variety of experimental and observational probes. These include proton-nucleus elastic scattering \cite{Zenihiro2018_Ca,Terashima2008_Sn,Zenihiro2010_Pb}, parity-violating electron scattering \cite{PhysRevLett.129.042501,PhysRevLett.126.172502}, dipole response measurements such as giant and pygmy dipole resonances \cite{Klimkiewicz2007_PDR,Carbone2010_PDR,Wieland2009_PDR,Tamii2011_alphaD,RocaMaza2013_alphaD,Paar2007_Review,Reinhard2010_Skin}, anti-analog giant dipole resonances \cite{Zenihiro2010_Pb,Yasuda2013AGDR,krasznahorkay2013anti_analog,RocaMaza2015AGDR,Krmpotic2007AGDR,Paar2007_Review}, antiproton annihilation at the nuclear surface \cite{PhysRevC.76.014311,PhysRevLett.87.082501}, pionic atoms and pion-induced reactions  \cite{Friedman2012NeutronSkins,baa6f013921d432f91aa6ba3daa1db4e}, coherent pion photoproduction \cite{PhysRevLett.112.242502}, and, more recently, multimessenger observations of neutron stars \cite{PhysRevD.104.063032,PhysRevLett.127.192701}. Each of these approaches probes the neutron density through distinct physical mechanisms and relies on different theoretical mappings to infer $\Delta r_{np}$.

While these methods provide valuable and complementary information, they also introduce significant challenges for global interpretation. Many neutron-skin extractions rely on reaction models, energy-density functional (EDF) correlations, or equation-of-state mappings that introduce method-dependent systematic uncertainties. Consequently, simple averages or weighted least-squares combinations of published values implicitly assume that all measurements are unbiased estimators of the same underlying quantity and that quoted uncertainties fully capture all sources of error. Such assumptions can lead to overconfident constraints, artificial tensions among datasets, and biased inferences of isovector EoS parameters.

In this work, we introduce a hierarchical Bayesian framework that enables a statistically consistent synthesis of heterogeneous neutron-skin constraints. The neutron-skin thickness is modeled as a smooth latent function of isospin asymmetry and nuclear size, while method-dependent bias parameters and intrinsic nuisance widths are introduced to account for unmodeled theoretical and experimental systematics. This hierarchical structure allows the data themselves to determine the effective weight of each probe, preventing any single method from dominating the inference due to underestimated uncertainties.

We focus on the tin isotopic chain ($Z=50$), which provides an ideal testing ground due to its wide span in isospin asymmetry and the availability of diverse neutron-skin constraints. Posterior sampling of the hierarchical model yields posterior-predictive neutron-skin systematics along the entire Sn chain, with quantified uncertainties. These latent systematics are then used as a calibrated intermediary to assess the consistency of nuclear energy-density functionals and to constrain the symmetry-energy slope parameter $L$ in a controlled and transparent manner. The choice of the Sn isotopic chain is motivated by its broad coverage in isospin asymmetry and the availability of multiple independent neutron-skin measurements. While the present analysis focuses on Sn isotopes, the hierarchical framework is general and can be applied to other isotopic chains. However, chains with more limited isospin variation or fewer experimental constraints would provide reduced sensitivity to the density dependence of the symmetry energy.

The primary objective of this study is to establish a unified and statistically robust framework for interpreting neutron-skin data and for quantifying their implications for the isovector sector of the nuclear EoS at sub-saturation densities. By propagating experimental, theoretical, and model-class uncertainties in a Bayesian manner, we aim to clarify the empirical information content of neutron skins and to strengthen the quantitative connection between laboratory nuclear experiments and neutron-star phenomenology.

The remainder of this paper is organized as follows. Section~II describes the neutron-skin dataset and the hierarchical Bayesian model. Section~III presents posterior inferences for the Sn isotopic chain and discusses the resulting uncertainty bands. Sections IV and V explore the implications of these results for the symmetry-energy slope parameter $L$ and compares them with constraints from astrophysical observations. Finally, Section~VI summarizes our conclusions and outlines prospects for future work.

\section{Experimental and Observational Probes of the Neutron Skin}

Neutron-skin thicknesses are not direct observables but are inferred from experimental or observational data through theoretical models that relate measured quantities to neutron and proton density distributions. Consequently, the quoted uncertainties associated with neutron-skin extractions often include significant model-dependent components in addition to experimental statistical errors. Different probes access the neutron density through distinct physical mechanisms and theoretical mappings, leading to heterogeneous sources of systematic uncertainty that must be treated carefully in any global analysis.

A central objective of this work is to incorporate these diverse constraints in an internally consistent manner. Different groups/frameworks can shift central values at the $\approx 0.02-0.05$ fm level even using similar data. However, when compiling a larger meta-table, it is best to track (i) beam energy, (ii) reaction framework, (iii) density parametrization freedom, (iv) whether the uncertainty is experimental-only or experimental plus model.  Rather than assuming that all neutron-skin measurements are unbiased estimators of a single underlying quantity with fully characterized uncertainties, we explicitly account for method-dependent systematics within a hierarchical Bayesian framework, as described in Sec.~III. Below we briefly summarize the physical content and dominant uncertainties associated with the different probes included in this study.  In our data collection multiple entries for, e.g., $^{132}$Sn or $^{208}$Pb are treated as features, i.e., independent inputs.

\subsection{Hadronic Probes}

Proton-nucleus elastic scattering at intermediate energies provides sensitivity to neutron densities through the isovector component of the nucleon-nucleus optical potential. This method benefits from relatively high precision and broad coverage across isotopic chains. However, the extraction of neutron densities relies on reaction models, typically based on relativistic impulse approximations with effective nucleon-nucleon interactions, which introduce model-dependent uncertainties. Variations in beam energy, optical potential parametrizations, and density functional assumptions can shift inferred neutron-skin values at the level of $\approx 0.02$-$0.05$~fm, even when analyzing similar datasets. In the present work, neutron-skin constraints from proton elastic scattering are taken from Refs.~\cite{Zenihiro2010_Pb,Zenihiro2018_Ca,Terashima2008_Sn}.

Antiproton annihilation at the nuclear surface provides a complementary hadronic probe with enhanced sensitivity to surface neutrons. Slow antiprotons form antiprotonic atoms and annihilate predominantly in the nuclear periphery, yielding constraints on neutron density distributions with comparatively modest model dependence. While uncertainties associated with annihilation dynamics and density-shape assumptions remain, antiprotonic atoms are among the cleanest probes of surface neutron distributions. Neutron-skin constraints from antiprotonic atoms are taken from Refs.~\cite{PhysRevC.76.014311,PhysRevLett.87.082501}.

\subsection{Electromagnetic Probes}
Parity-violating electron scattering (PVES) provides a theoretically clean probe of neutron densities via the weak interaction. Measurements such as PREX and CREX~\cite{PhysRevLett.126.172502,PhysRevLett.129.042501} directly access the weak form factor of nuclei and are largely free of strong-interaction uncertainties. Their current limitations are primarily experimental, including small asymmetries, limited statistics, and the restricted number of measured nuclei. In this work we adopt the published PVES neutron-skin values for $^{48}$Ca and $^{208}$Pb~\cite{PhysRevLett.126.172502,PhysRevLett.129.042501}, treating the quoted uncertainties as combined experimental and theoretical contributions.

In collisions of ultrarelativistic nuclei, the intense Lorentz-contracted electromagnetic field of a fast-moving ion can be treated as a flux of linearly polarized quasi-real photons \cite{BERTULANI1988299}. When two heavy ions pass each other at impact parameters exceeding the sum of their nuclear radii, these photons may fluctuate into virtual quark-antiquark pairs that interact with the gluon field of the opposing nucleus, producing vector mesons such as the $\rho^0$. In Ref. \cite{doi:10.1126/sciadv.abq3903}, the photon polarization was exploited in diffractive photoproduction, revealing a characteristic spin-dependent interference structure in the angular distribution of the $\rho^0 \rightarrow \pi^+\pi^-$ decay channel.
The interference pattern arises from the coherent superposition of production amplitudes separated by distances far exceeding the $\rho^0$ propagation length within its lifetime. Analysis of the diffractive scattering data allowed extraction of the effective strong-interaction radii, both exceeding their corresponding charge radii \cite{doi:10.1126/sciadv.abq3903}.

Dipole response measurements, including the giant dipole resonance (GDR), pygmy dipole resonance (PDR), and anti-analog giant dipole resonance (AGDR), constrain neutron skins indirectly through correlations with the symmetry energy. These observables probe collective isovector excitations and are typically interpreted using energy-density functional or random-phase approximation systematics. As a result, their uncertainties are dominated by correlation and model dependence rather than by experimental resolution alone. While such probes benefit from extensive datasets across many nuclei, different EDF calibrations and correlation choices can lead to variations in inferred neutron-skin values at the few $10^{-2}$~fm level. Neutron-skin constraints derived from dipole response measurements are taken from Refs.~\cite{Klimkiewicz2007_PDR,Carbone2010_PDR,Wieland2009_PDR,Tamii2011_alphaD,RocaMaza2013_alphaD,Paar2007_Review,Reinhard2010_Skin,Zenihiro2010_Pb,Yasuda2013AGDR,krasznahorkay2013anti_analog,RocaMaza2015AGDR,Krmpotic2007AGDR,Paar2007_Review} and related works. The sensitivity to neutron skins in GDR measurements arises because
\begin{equation}
\alpha_D = \frac{A\left< r^2 \right>}{J} \left( 1 + \kappa {L\over J} \right),
\end{equation}
which induces a correlation with $\Delta r_{np}$. GDR observables tend to be smoother because they probe the entire nuclear volume, not just the surface. As a result, GDR-inferred skins also have larger model uncertainties, are less sensitive to shell structure, and tend to give smaller spreads than PDR-based values.

\subsection{Mesonic Probes}

Pionic atoms and pion-induced reactions probe neutron densities near the nuclear surface through strong-interaction effects. Analyses of pionic atoms rely on optical potential models and assumptions about neutron density shapes, leading to nontrivial systematic uncertainties. Complementary information is provided by $\pi^+$ total reaction cross sections, which probe slightly higher densities and exhibit different sensitivities to Coulomb effects and pion-nucleus interactions. These probes provide valuable cross-validation of surface neutron distributions but must be treated with care due to their model dependence. Neutron-skin constraints from pionic probes are taken from Refs.~\cite{Friedman2012NeutronSkins}.

The $\pi^+$ total reaction cross section constraints are the natural companion to pionic atoms, and it is good to keep them explicitly separate (they probe a slightly deeper surface density and have different systematics). Usually, one measures the  total reaction cross sections  $\sigma_R(\pi^+)$ on  nuclei at low/intermediate energies. Then the data are fitted with an optical potential calculation constrained by $\pi N$ scattering. The neutron-skin sensitivity arises because $\pi^+$ interacts more weakly with neutrons than $\pi^-$. Compared to pionic atoms it probes slightly higher densities, has smaller Coulomb distortions, and gives an independent handle on neutron-skin thickness. The data sensitivity sits between pionic atoms (very low-density surface) and hadronic probes (p and $\bar p$). Systematics are different from $\pi$ atoms allowing for a great  cross-validation of $\Delta r_{np}$.   In Ref. \cite{Friedman2012NeutronSkins} $\pi^+$ total reaction cross sections used to infer a neutron skin are from an analysis of 
the experiment in Ref. \cite{baa6f013921d432f91aa6ba3daa1db4e}.) Coherent  $\pi^0$  photoproduction on nuclei also allows a standard neutron-skin extraction. Information on the size and shape of the neutron skin on  $^{208}$Pb has extracted from coherent pion photoproduction \cite{PhysRevLett.112.242502}.

\subsection{Astrophysical Constraints}
Neutron-star observations, including tidal deformabilities from gravitational-wave events and radius measurements from X-ray timing, constrain the symmetry energy and thus neutron skins indirectly. These constraints probe matter at densities beyond those in finite nuclei and require an equation-of-state mapping to laboratory observables. While powerful, they must be incorporated carefully to avoid double counting correlated information.

\begin{figure}[t]
\includegraphics[width=8.7cm]{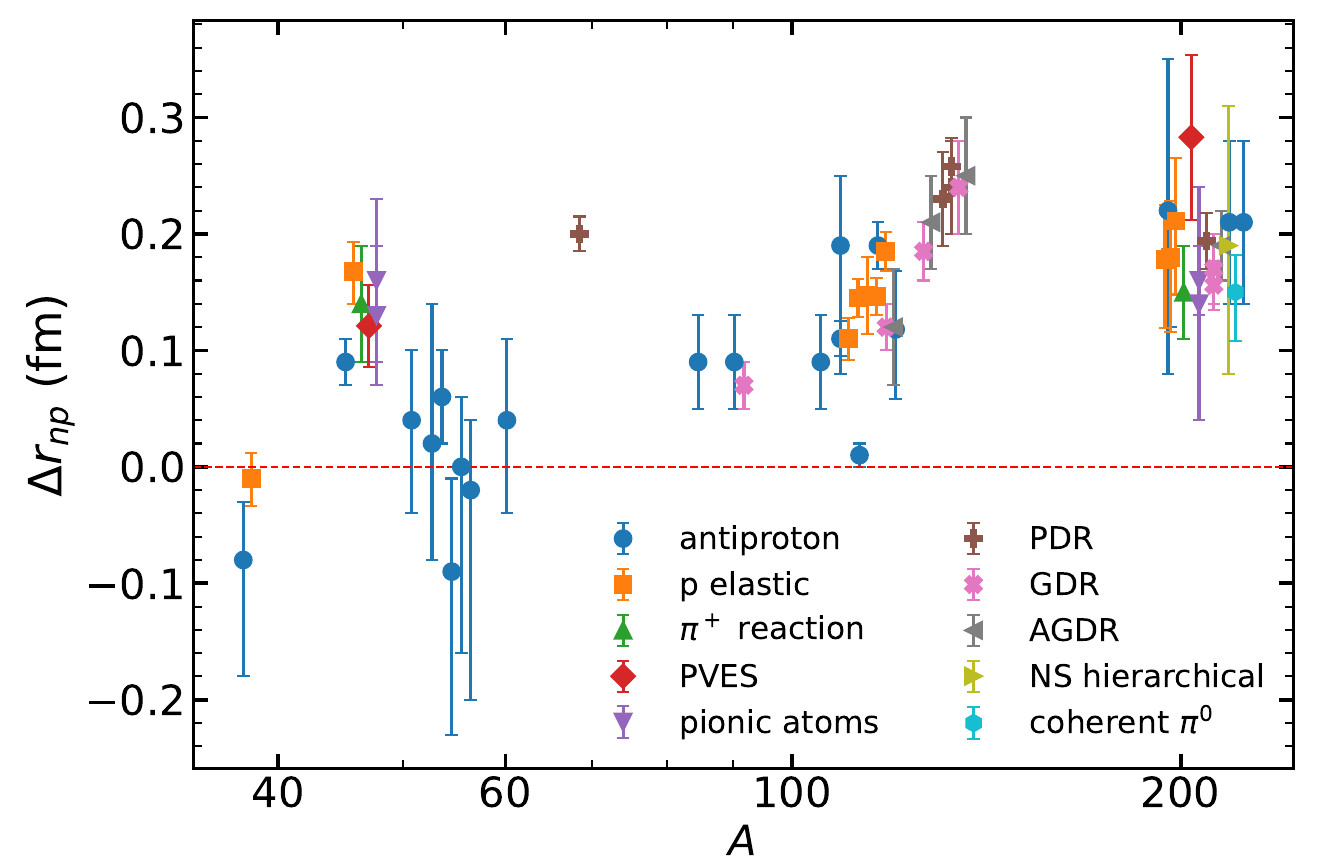}
\caption{Neutron-skin data collected from several recent sources, including antiproton annihilation, proton elastic scattering, PVES, pionic atoms, PDR, GDR, AGDR, coherent pion production,  relativistic heavy ions, and neutron stars observations. See text for details.}
\label{skindata}
\end{figure}

Neutron-star tidal deformability observations (primarily GW170817 tidal deformability, plus in some analyses heavy pulsar masses and NICER) need an EoS inference and then a mapping from astrophysical constraints to the symmetry-energy slope $L \rightarrow \Delta r_{np}$ ($^{208}$Pb). Therefore, the ``error bars" reflect both observational and modeling/mapping uncertainties. The neutron-skin constraints mapped from NICER neutron-star observations are usually inferred from Bayesian-inference works \cite{PhysRevD.104.063032,PhysRevLett.127.192701}.

In Table \ref{tabdata} we list the 59 neutron-skin values extracted from numerous experiments, observations,  and theoretical analyses.  In Fig. \ref{skindata} we plot the neutron-skin data collected, which have been used in our statistical analysis.

\begin{table*}[t]
\caption{Experimental and observational neutron-skin thickness data used in this work.
Shown are the nucleus, probe type, extracted neutron skin $\Delta r_{np}$, and
the corresponding references. Asymmetric uncertainties are given when available.
Abbreviations for probe types are defined in the text.}
\label{tabdata}
\centering
\renewcommand{\arraystretch}{1.10}
\setlength{\tabcolsep}{4pt}

\begin{tabular}{l l c p{3.2cm} @{\hspace{0.8cm}} l l c p{3.2cm}}
\hline\hline
Nucleus & Probe & $\Delta r_{np}$ (fm) & Ref. &
Nucleus & Probe & $\Delta r_{np}$ (fm) & Ref. \\
\hline

$^{40}$Ca  & $p$-elast. & $-0.010^{+0.022}_{-0.024}$ & \cite{Zenihiro2018_Ca} &
$^{68}$Ni  & PDR & $0.200\pm0.015$ & \cite{Klimkiewicz2007_PDR,Carbone2010_PDR,Wieland2009_PDR,Tamii2011_alphaD,RocaMaza2013_alphaD,Paar2007_Review,Reinhard2010_Skin} \\

$^{48}$Ca  & $p$-elast. & $0.168^{+0.025}_{-0.028}$ & \cite{Zenihiro2018_Ca} &
$^{130}$Sn & PDR & $0.230\pm0.040$ & \cite{Klimkiewicz2007_PDR,Carbone2010_PDR,Wieland2009_PDR,Tamii2011_alphaD,RocaMaza2013_alphaD,Paar2007_Review,Reinhard2010_Skin} \\

$^{116}$Sn & $p$-elast. & $0.110\pm0.018$ & \cite{Terashima2008_Sn} &
$^{132}$Sn & PDR & $0.258\pm0.024$ & \cite{Klimkiewicz2007_PDR,Carbone2010_PDR,Wieland2009_PDR,Tamii2011_alphaD,RocaMaza2013_alphaD,Paar2007_Review,Reinhard2010_Skin} \\

$^{118}$Sn & $p$-elast. & $0.145\pm0.016$ & \cite{Terashima2008_Sn} &
$^{208}$Pb & PDR & $0.194\pm0.024$ & \cite{Klimkiewicz2007_PDR,Carbone2010_PDR,Wieland2009_PDR,Tamii2011_alphaD,RocaMaza2013_alphaD,Paar2007_Review,Reinhard2010_Skin} \\

$^{120}$Sn & $p$-elast. & $0.147\pm0.033$ & \cite{Terashima2008_Sn} &
$^{90}$Zr  & GDR & $0.070\pm0.020$ & \cite{Klimkiewicz2007_PDR,Carbone2010_PDR,Wieland2009_PDR,Tamii2011_alphaD,RocaMaza2013_alphaD,Paar2007_Review,Reinhard2010_Skin} \\

$^{122}$Sn & $p$-elast. & $0.146\pm0.016$ & \cite{Terashima2008_Sn} &
$^{116}$Sn & GDR & $0.120\pm0.020$ & \cite{Klimkiewicz2007_PDR,Carbone2010_PDR,Wieland2009_PDR,Tamii2011_alphaD,RocaMaza2013_alphaD,Paar2007_Review,Reinhard2010_Skin} \\

$^{124}$Sn & $p$-elast. & $0.185\pm0.017$ & \cite{Terashima2008_Sn} &
$^{124}$Sn & GDR & $0.185\pm0.025$ & \cite{Klimkiewicz2007_PDR,Carbone2010_PDR,Wieland2009_PDR,Tamii2011_alphaD,RocaMaza2013_alphaD,Paar2007_Review,Reinhard2010_Skin} \\

$^{204}$Pb & $p$-elast. & $0.178^{+0.047}_{-0.059}$ & \cite{Zenihiro2010_Pb} &
$^{132}$Sn & GDR & $0.240\pm0.040$ & \cite{Klimkiewicz2007_PDR,Carbone2010_PDR,Wieland2009_PDR,Tamii2011_alphaD,RocaMaza2013_alphaD,Paar2007_Review,Reinhard2010_Skin} \\

$^{206}$Pb & $p$-elast. & $0.180^{+0.048}_{-0.064}$ & \cite{Zenihiro2010_Pb} &
$^{208}$Pb & GDR & $0.156\pm0.021$ & \cite{Klimkiewicz2007_PDR,Carbone2010_PDR,Wieland2009_PDR,Tamii2011_alphaD,RocaMaza2013_alphaD,Paar2007_Review,Reinhard2010_Skin} \\

$^{208}$Pb & $p$-elast. & $0.211^{+0.054}_{-0.063}$ & \cite{Zenihiro2010_Pb} &
$^{116}$Sn & AGDR & $0.120\pm0.050$ & \cite{Yasuda2013AGDR,krasznahorkay2013anti_analog,RocaMaza2015AGDR,Krmpotic2007AGDR,Paar2007_Review} \\

$^{48}$Ca  & $\bar p$ & $0.130\pm0.030$ & \cite{PhysRevC.76.014311,PhysRevLett.87.082501} &
$^{124}$Sn & AGDR & $0.210\pm0.040$ & \cite{Yasuda2013AGDR,krasznahorkay2013anti_analog,RocaMaza2015AGDR,Krmpotic2007AGDR,Paar2007_Review} \\

$^{90}$Zr  & $\bar p$ & $0.090\pm0.030$ & \cite{PhysRevC.76.014311,PhysRevLett.87.082501} &
$^{132}$Sn & AGDR & $0.250\pm0.050$ & \cite{Yasuda2013AGDR,krasznahorkay2013anti_analog,RocaMaza2015AGDR,Krmpotic2007AGDR,Paar2007_Review} \\

$^{116}$Sn & $\bar p$ & $0.110\pm0.030$ & \cite{PhysRevC.76.014311,PhysRevLett.87.082501} &
$^{208}$Pb & AGDR & $0.190\pm0.030$ & \cite{Yasuda2013AGDR,krasznahorkay2013anti_analog,RocaMaza2015AGDR,Krmpotic2007AGDR,Paar2007_Review} \\

$^{124}$Sn & $\bar p$ & $0.180\pm0.030$ & \cite{PhysRevC.76.014311,PhysRevLett.87.082501} &
$^{48}$Ca  & PVES & $0.121\pm0.035$ & \cite{PhysRevLett.129.042501} \\

$^{204}$Pb & $\bar p$ & $0.170\pm0.040$ & \cite{PhysRevC.76.014311,PhysRevLett.87.082501} &
$^{208}$Pb & PVES & $0.283\pm0.071$ & \cite{PhysRevLett.126.172502} \\

$^{206}$Pb & $\bar p$ & $0.180\pm0.040$ & \cite{PhysRevC.76.014311,PhysRevLett.87.082501} &
$^{48}$Ca  & $\pi^-$ atom & $0.130\pm0.060$ & \cite{Friedman2012NeutronSkins,baa6f013921d432f91aa6ba3daa1db4e} \\

$^{208}$Pb & $\bar p$ & $0.200\pm0.040$ & \cite{PhysRevC.76.014311,PhysRevLett.87.082501} &
$^{208}$Pb & $\pi^-$ atom & $0.150\pm0.080$ & \cite{Friedman2012NeutronSkins,baa6f013921d432f91aa6ba3daa1db4e} \\

$^{48}$Ca  & $\sigma(\pi^+)$ & $0.140\pm0.050$ & \cite{Friedman2012NeutronSkins,baa6f013921d432f91aa6ba3daa1db4e} &
$^{208}$Pb & coh.\ $\pi^0$ & $0.150^{+0.032}_{-0.042}$ & \cite{PhysRevLett.112.242502} \\

$^{90}$Zr  & $\sigma(\pi^+)$ & $0.090\pm0.050$ & \cite{Friedman2012NeutronSkins,baa6f013921d432f91aa6ba3daa1db4e} &
$^{208}$Pb & NS & $0.17\pm0.04$ & \cite{PhysRevLett.127.192701} \\

$^{116}$Sn & $\sigma(\pi^+)$ & $0.120\pm0.050$ & \cite{Friedman2012NeutronSkins,baa6f013921d432f91aa6ba3daa1db4e} &
$^{208}$Pb & NS & $0.204^{+0.030}_{-0.026}$ & \cite{PhysRevD.104.063032} \\

$^{208}$Pb & $\sigma(\pi^+)$ & $0.150\pm0.040$ & \cite{Friedman2012NeutronSkins,baa6f013921d432f91aa6ba3daa1db4e} &
$^{197}$Au & Rel. ions & $0.17\pm0.08$ & \cite{doi:10.1126/sciadv.abq3903} \\

 & & & &
$^{238}$U & Rel. ions & $0.44\pm0.08$ & \cite{doi:10.1126/sciadv.abq3903} \\

\hline\hline
\end{tabular}
\end{table*}

\section{Hierarchical Bayesian Framework}

A common approach in the literature is to combine neutron-skin extractions from different probes using weighted averages or least-squares fits based on quoted uncertainties. Such procedures implicitly assume that all measurements are unbiased estimators of the same underlying quantity and that their reported uncertainties fully capture all sources of error. Given the heterogeneous nature of neutron-skin probes, these assumptions are generally not satisfied and can lead to overconfident inferences and artificial tensions among datasets.

To address these issues, we adopt a hierarchical Bayesian framework in which the neutron-skin thickness of each nucleus is treated as a latent quantity governed by a smooth, physically motivated trend, while method-dependent systematics are modeled explicitly. This structure allows the data themselves to determine the effective weight of each probe and ensures self-consistency within the model propagation of experimental and theoretical uncertainties.

\subsection{Latent Neutron-Skin Model}
The neutron-skin thickness is a finite-size observable that arises from the interplay between bulk symmetry energy, surface tension, and Coulomb effects. Guided by droplet-model considerations and geometric scaling arguments, we model the latent neutron-skin thickness as a function of isospin asymmetry and nuclear size,
\begin{equation}
\Delta r_{np}(A,Z,N) =
\beta_0
+ \beta_1 I
+ \beta_2 A^{1/3}
+ \beta_3 I A^{1/3},
\label{eq:latent_model}
\end{equation}
where $I = (N-Z)/A$ is the isospin asymmetry. This parametrization captures the leading dependence on bulk asymmetry and surface geometry while remaining sufficiently flexible for global inference.

The $I$-dependent terms encode the response of the system to symmetry pressure, which is strongly correlated with the slope of the symmetry energy, whereas the $A^{1/3}$ terms reflect finite-size and Coulomb effects associated with rms-radius definitions. The mixed term $I A^{1/3}$ is particularly important for describing isotopic trends in heavy nuclei, where surface-symmetry effects play a dominant role. Higher-order terms are not included, as the available data do not support additional complexity. Equation \eqref{eq:latent_model} should be viewed as a leading-order parametrization of neutron-skin systematics. Possible higher-order contributions are effectively absorbed by the hierarchical model through nuisance parameters, and we have verified that their inclusion does not significantly modify the inferred symmetry-energy constraints (see Appendix).

The appearance of $A^{1/3}$ terms also follows directly from the nuclear droplet model and its modern refinements. In its standard formulation, the neutron-skin thickness can be written schematically as \cite{Myers1969,Myers1974,Danielewicz2003,Centelles2009,RocaMaza2011}
\begin{equation}
\Delta r_{np}
\approx
\frac{2 r_0}{3}
\frac{J}{Q}
\frac{I - I_C}{1 + \frac{9J}{4Q} A^{-1/3}}  \Delta r_{\rm Coul},
  \end{equation}
  where $J$ is the symmetry energy at saturation, $Q$ is the surface stiffness coefficient, $I_C$ is a Coulomb correction, and $\Delta r_{\rm Coul}$ denotes additional Coulomb contributions.

Expanding the denominator for large $A$ yields
\begin{equation}
\Delta r_{np}
\approx
\alpha I + \beta I A^{-1/3} + \gamma A^{1/3} + \cdots. \label{globtrend}
\end{equation}
 When expressed in terms of rms radii, which themselves scale with $R \propto A^{1/3}$, the observable neutron skin acquires effective contributions proportional to $A^{1/3}$ and $I A^{1/3}$.
Thus, 
$\beta_1 I$ represents the bulk symmetry-energy contribution, strongly correlated with the slope parameter $L$ of the symmetry energy.
$\beta_3 I A^{1/3}$ encodes the surface symmetry energy and the surface-to-volume competition. This term dominates isotopic trends and is essential for describing heavy, neutron-rich nuclei.
$\beta_2 A^{1/3}$ primarily captures Coulomb geometry and systematic effects associated with rms-radius definitions.
 $\beta_0$ absorbs small offsets arising from higher-order corrections and residual model dependencies.

For the tin isotopic chain ($Z=50$), $A^{1/3}$ varies by approximately 15\% between $^{100}$Sn and $^{138}$Sn, while the asymmetry $I$ spans the range from nearly zero to $I \approx eq 0.28$. The dominant growth of the neutron skin along the chain is therefore governed by the combined dependence
\begin{equation}
\Delta r_{np}^{\rm Sn}(A) \approx \beta_1 I(A) + \beta_3 I(A) A^{1/3}. \label{DRL}
\end{equation}
Omitting the $A^{1/3}$ dependence would lead to an artificially linear behavior in $I$ and systematically underestimate neutron skins in the heavier isotopes, as well as their associated uncertainties.

Each neutron-skin measurement $y_i$ is related to the corresponding latent value through
\begin{equation}
y_i = \Delta r_{np}(A_i,Z_i,N_i) + b_{m(i)} + \epsilon_i,
\end{equation}
where $b_{m(i)}$ is a method-dependent bias parameter associated with probe $m$, and $\epsilon_i$ represents residual fluctuations. The bias parameters account for systematic offsets in the mapping from experimental observables to neutron-skin thicknesses and shift the mean prediction of a given method relative to the reference scale. The parameters $b_m$ and $tau_m$ allow for method-dependent biases and intrinsic scatter beyond quoted uncertainties. These parameters are inferred simultaneously with the global neutron-skin trend, enabling a statistically consistent propagation of systematic uncertainties across different experimental probes.

The residuals are modeled as Gaussian,
\begin{equation}
\epsilon_i \approx \mathcal{N}\!\left(0,\;\sigma_{i,\mathrm{exp}}^2 + \tau_{m(i)}^2\right),
\end{equation}
where $\sigma_{i,\mathrm{exp}}$ is the quoted experimental uncertainty and $\tau_m$ is an intrinsic nuisance width specific to method $m$. The nuisance widths implement variance inflation and account for unmodeled theoretical uncertainties, missing correlations, and underreported systematic effects. Unlike bias parameters, $\tau_m$ broadens the likelihood without shifting its mean, thereby reducing the influence of overconfident but model-dependent probes.

We adopt weakly informative priors on all model parameters to regularize inference while allowing the data to dominate the posterior. The global trend parameters $\{\beta_k\}$ are assigned broad Gaussian priors, while method-dependent bias parameters $b_m$ are centered at zero with widths reflecting prior uncertainty about systematic offsets. Nuisance widths $\tau_m$ are constrained to be positive and are assigned half-normal priors with scales chosen to be non-restrictive relative to expected neutron-skin variations.

The joint posterior distribution of all parameters is sampled using an affine-invariant ensemble Markov Chain Monte Carlo algorithm \cite{GoodmanWeare2010,ForemanMackey2013}. This approach efficiently explores high-dimensional and correlated parameter spaces without requiring fine-tuning of proposal distributions. Convergence is assessed using standard diagnostics, and posterior samples are used to generate posterior-predictive distributions for neutron-skin thicknesses of individual nuclei and isotopic chains. This procedure naturally propagates experimental uncertainties, method-dependent systematics, and model uncertainty into the predicted neutron-skin bands. The resulting posterior-predictive distributions provide a transparent representation of the empirical information content of the neutron-skin dataset and form the basis for subsequent comparisons with nuclear energy-density functionals.

\section{Results and Implications for the Symmetry Energy}

\subsection{Method-Dependent Offsets and Intrinsic Scatter}

We summarize the posterior distributions of the ``bias" $b_m$ and the inferred nuisance widths $\tau_m$ in Table~\ref{tab:method_systematics}. $b_m$  captures method-specific calibration shifts in the mapping from the raw observable to $\Delta r_{np}$, e.g., optical potential choices for hadronic scattering, EDF correlation calibration for dipole observables, pionnucleus optical potential for pionic atoms, etc. It  is defined relative to a reference method. One method is ``pinned" to $b_m=0$, by construction (here, proton-elastic scattering).
Therefore, other methods are ``relative to the pinned method, preferring skins shifted by $b_m$".
A positive sign means that this method tends to infer larger neutron skins than the reference, for the same underlying nuclei/trend, whereas a negative sign tends to infer smaller skins than the reference.
A $|b_m| \approx 0.01$ fm is tiny, often below practical sensitivity, $|b_m| \approx 0.03-0.05$ fm is already a meaningful systematic shift in neutron-skin physics, and $|b_m| \gtrsim 0.07$ fm is large, suggesting either strong model dependence, hidden correlations, or tensions with the global trend model. 

\begin{table}[t]
\caption{Posterior summaries of method-dependent systematics: bias $b_m$ and intrinsic nuisance width $\tau_m$. Uncertainties correspond to 68\% credible intervals.}
\label{tab:method_systematics}
\begin{ruledtabular}
\begin{tabular}{l c c}
Method & $b_m$ (fm) & $\tau_m$ (fm) \\
\hline
Proton elastic scattering & 0 & 0.014 $\pm$ 0.007 \\
Dipole response (AGDR) & 0.014 $\pm$ 0.025 & 0.022$^{+0.016}_{-0.010}$ \\
Dipole response (GDR) & -0.008 $\pm$ 0.016 & 0.018 $\pm$ 0.009 \\
Astrophysical (GW+NICER) & 0.001 $\pm$ 0.066 & 0.029$^{+0.027}_{-0.014}$ \\
Dipole response (PDR) & 0.021 $\pm$ 0.019 & 0.018 $\pm$ 0.010 \\
Parity-viol. $e$ scatt. (PVES) & -0.005 $\pm$ 0.039 & 0.032$^{+0.028}_{-0.016}$ \\
Antiproton annihilation & -0.002 $\pm$ 0.016 & 0.016 $\pm$ 0.009 \\
Coherent $\pi^0$ photoproduction & -0.028 $\pm$ 0.044 & 0.028$^{+0.026}_{-0.014}$ \\
$\pi^+$ total reaction & -0.014 $\pm$ 0.027 & 0.023$^{+0.018}_{-0.010}$ \\
Pionic atoms & -0.022 $\pm$ 0.038 & 0.024$^{+0.020}_{-0.011}$ \\
\end{tabular}
\end{ruledtabular}
\end{table}

We read from Table~\ref{tab:method_systematics} that the AGDR (model)  has slight preference for ``larger" skins, GDR is essentially unbiased, PDR has a moderately positive shift,   PVES is fully consistent with zero, antiproton has no systematic offset,   and astrophysical (GW+NICER) has a poorly constrained offset. There is no statistically significant global bias for any method.
All biases are consistent with zero at the $1 \sigma-2\sigma$ level. This means that apparent tensions between probes are not due to large systematic shifts, but rather to differences in scatter and model dependence, captured by $\tau_m$. 

The nuisance widths $\tau_m$ encode additional variance beyond the quoted experimental uncertainties and therefore quantify unmodeled theoretical and systematic effects. Hadronic probes and dipole-response observables exhibit moderate nuisance widths, reflecting uncertainties associated with reaction models and energy-density functional correlations. Parity-violating electron scattering shows a comparatively larger $\tau_m$, which primarily reflects limited statistics and the small number of measured nuclei rather than theoretical deficiencies. Astrophysical constraints display the largest nuisance widths, as expected from the indirect mapping between neutron-star observables and finite-nucleus properties. Within the hierarchical framework, probes with larger $\tau_m$ are automatically down-weighted through variance inflation, preventing overconfident datasets from dominating the global inference.

\subsection{Global neutron-skin trend parameters}

Posterior summaries of the global trend parameters in Eq.~(\ref{eq:latent_model}) are given in Table~\ref{tab:global_trend}. The inferred coefficients describe a smooth neutron-skin systematics dominated by isospin asymmetry, with finite-size and surface-symmetry effects providing important corrections for heavy nuclei. The mixed term proportional to $I A^{1/3}$ plays a central role in reproducing the curvature of isotopic trends and would be poorly constrained in single-nucleus analyses.

\begin{figure}[t]
\includegraphics[width=8.8cm]{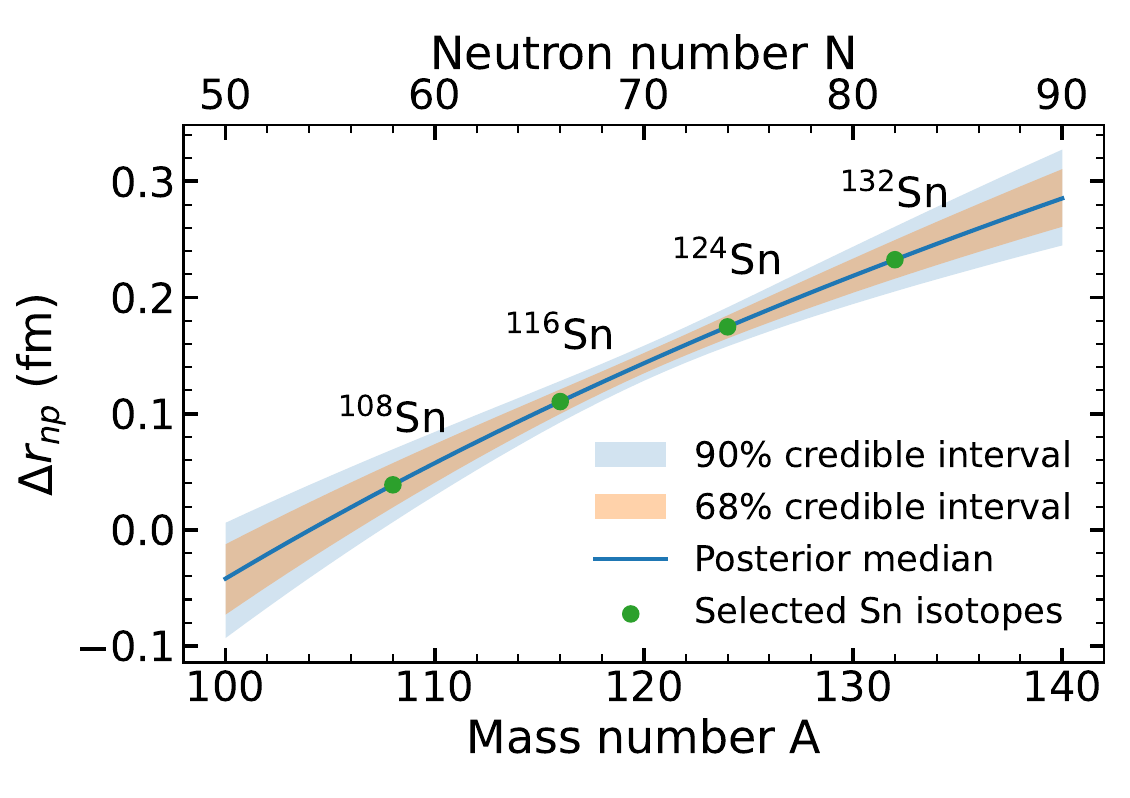}
\caption{Posterior-predictive neutron-skin thickness $\Delta r_{np}$ as a function of neutron number $N$ along the tin isotopic chain. The solid line denotes the posterior median, while the shaded bands indicate 68\% and 90\% credible intervals.}
\label{fig:sn_chain}
\end{figure}

Fig.~\ref{fig:sn_chain} displays the posterior-predictive neutron-skin thickness along the tin isotopic chain. The median prediction increases monotonically with neutron number, reflecting the growing symmetry pressure in neutron-rich systems. The associated credible intervals exhibit a characteristic structure: uncertainties are smallest near $A\approx 120$, where the data density is highest and the latent model interpolates between well-constrained regions of parameter space, and increase toward both the proton-rich and neutron-rich ends of the chain, where extrapolation in isospin asymmetry becomes unavoidable. This behavior arises naturally from the geometry of the latent model and the covariance structure of the inferred parameters. 

\begin{table}[t]
\caption{Posterior summaries of the global neutron-skin trend parameters defined in Eq.~(\ref{eq:latent_model}). Uncertainties correspond to 68\% credible intervals.}
\label{tab:global_trend}
\begin{ruledtabular}
\begin{tabular}{l c}
Parameter & Value (fm) \\
\hline
$\beta_0$ & 0.143 $\pm$ 0.080 \\
$\beta_1$ & 0.424 $\pm$ 0.348 \\
$\beta_2$  & -0.039 $\pm$ 0.021 \\
$\beta_3$ & 0.150 $\pm$ 0.089 \\
\end{tabular}
\end{ruledtabular}
\end{table}

The combined dataset includes multiple constraints in/near the stable Sn region. 
In a Bayesian regression, posterior variance is smallest where the design matrix has the most support, i.e., where one has 
the most informative constraints at similar $I$ and similar $A^{1/3}$.
For Sn isotopes near $A \approx 120$, the values of $(I,A^{1/3})$
sit near the ``center of mass" of the training data cloud (not just Sn, but the whole dataset: Ca, Zr, Pb, etc.). 
That minimizes parameter covariance amplification.
The uncertainties increase toward both ends (light and heavy Sn). This is because we are extrapolating in isospin asymmetry
$I$.  For $^{100}$Sn $I$ is small and the skin is expected to be small, possibly even slightly negative within a purely statistical regression.
For large $N$, e.g., $^{136}$Sn, even if $A^{1/3}$ changes slowly, the lever arm in $I$  is large. With the parametric form of Eq. \eqref{globtrend},  the uncertainty in the slope is dictated by $\beta_1 + \beta_3 A^{1/3}$ which increases with $A$.  

Selected predictions of the latent model include
\begin{align}
\Delta r_{np}(^{124}\mathrm{Sn}) &= \text{0.1746}^{+\text{0.00964}}_{-\text{0.00959}}\ \mathrm{fm},\nonumber \\
\Delta r_{np}(^{132}\mathrm{Sn}) &= \text{0.2326}^{+\text{0.0122}}_{-\text{0.0161}}\ \mathrm{fm}.
\end{align}

\section{EDF bridge: mapping $\Delta r_{np}(N)$ to $(J,L)$}
Using an isotopic chain provides leverage in the isospin-asymmetry variable $I$
and thus constrains not only an overall offset but also the slope of $\Delta r_{np}$ versus $N$ (or $I$). This reduces degeneracies with surface terms and enhances discrimination among Energy Density Functionals (EDFs). Empirically, within common EDF families, the correlation between $\Delta r_{np}$ in neutron-rich Sn isotopes and $L$ is particularly strong.  Many EDFs show a 
$J$-$L$ correlation, which means skin data can indirectly restrict $J$, but typically with larger degeneracy. The present analysis employs a set of non-relativistic EDFs to ensure internal consistency and controlled variation of model parameters. Inclusion of relativistic models would broaden the theoretical uncertainty and may slightly shift the inferred symmetry-energy parameters, but is not expected to alter the qualitative conclusions. 

The EDF families we include in our calculations stem from Hartree-Fock-Bogoliubov (HFB) Skyrme interactions SIII \cite{BEINER197529}, SKA and SKB \cite{KOHLER1976301},  SKM* \cite{BARTEL198279}, SKP \cite{DOBACZEWSKI1984103}, UNE0 and UNE1 \cite{STOITSOV20131592}, SKMP \cite{Bennour1989}, SKI2, SKI3, SKI4 and SKI5 \cite{REINHARD1995467}, SLY230A \cite{Chabanat:97}, SLY4, SLY5, SLY6, and SLY7 \cite{CHABANAT1998231},  SKX \cite{Brown:1998}, SKO \cite{Reinhard:PRC.60.014316},  SK255 and SK272 \cite{Agrawal2003}, HFB9 \cite{GORIELY2005425} and SKXS20 \cite{Dutra:PRC.85.035201}.   
A mixed volume-surface pairing interaction is included, as described in Refs.  \cite{Bertsch:PRC79.034306,Bertulani.PRC.100.015802}.  The HFB calculations were done with a modified version of the HFBTO code \cite{STOITSOV20131592}. 
Each EDF in the ensemble supplies the values for the nuclear matter parameters $J$, $L$ and a predicted Sn-chain neutron-skin curve $\Delta r_{np}^{\rm EDF}(N)$.

We assume an inferred latent Sn-chain constraint in the form of a median curve and uncertainty band
\begin{equation}
\Delta r_{np}^{\rm latent}(N)\pm \sigma(N),
\end{equation}
where $\sigma(N)$ is taken from the 68\% credible interval of the posterior predictive distribution, e.g.
\begin{equation}
\sigma(N)\approx \frac{1}{2}\left[\Delta r_{np,\,{\rm hi68}}^{\rm latent}(N)-\Delta r_{np,\,{\rm low68}}^{\rm latent}(N)\right],
\end{equation}
where the indices  ``$low68$" (``$hi68$") are weighted 16-84\% quantiles at each $N$. 
We compute an EDF goodness-of-fit statistic against the latent constraint over the overlap set of neutron numbers $\mathcal{N}_{\rm overlap}$:
\begin{equation}
\chi^2_{\rm EDF}=\sum_{N\in\mathcal{N}_{\rm overlap}}
\frac{\left[\Delta r_{np}^{\rm EDF}(N)-\Delta r_{np}^{\rm latent}(N)\right]^2}{\sigma^2(N)}.
\label{eq:chi2}
\end{equation}
This yields EDF weights
\begin{equation}
w_{\rm EDF}\propto \exp\left(-\frac{1}{2}\chi^2_{\rm EDF}\right),
\qquad
\sum_{\rm EDF}w_{\rm EDF}=1,
\label{eq:weights}
\end{equation}
which define an EDF-weighted posterior over $(J,L)$:
\begin{equation}
p(J,L \mid \text{latent Sn}) \approx \sum_{\rm EDF} w_{\rm EDF}\,\delta(J-J_{\rm EDF})\,\delta(L-L_{\rm EDF}).
\end{equation}
From these weights we obtain weighted summaries (medians and credible intervals) and generate weighted posteriors $p(J)$ and $p(L)$ via weighted histograms.

\begin{figure}[t]
\includegraphics[width=8.8cm]{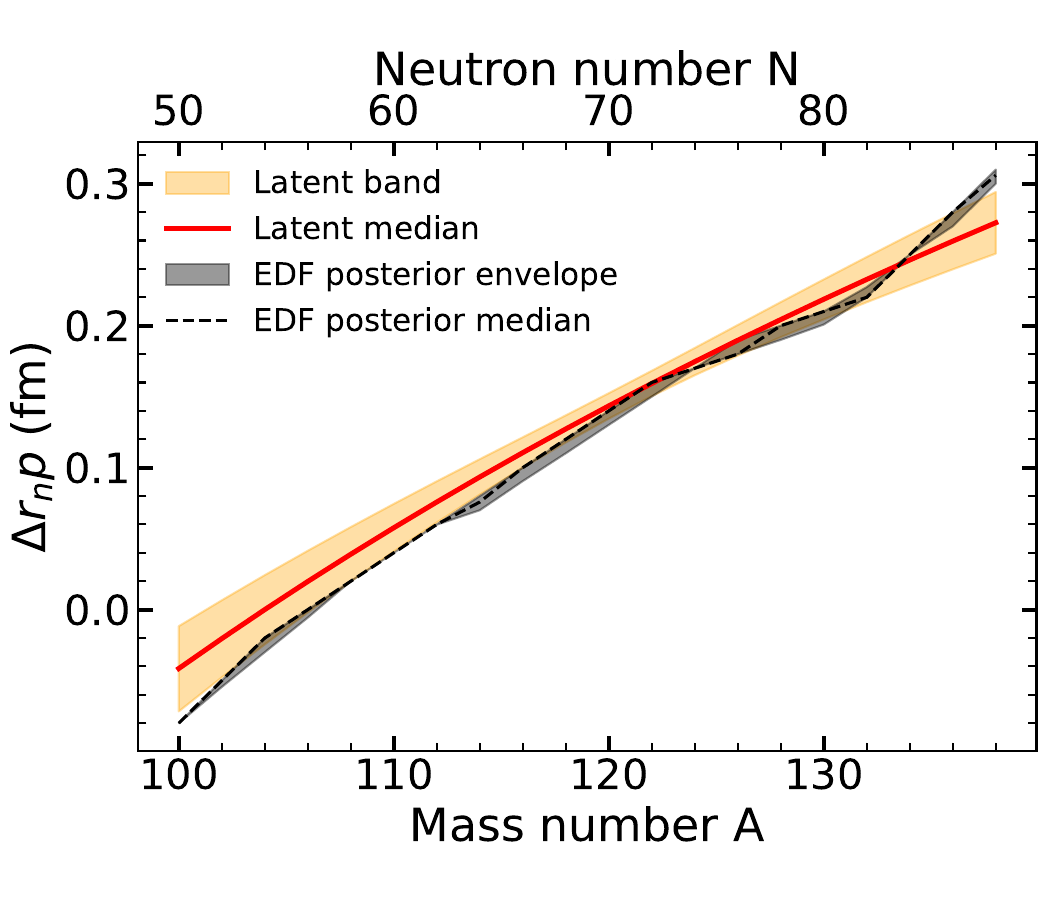}
\caption{The solid curve and shaded band show the inferred latent neutron-skin systematics along the Sn isotopic chain (same as Fig. \ref{fig:sn_chain}). The gray envelopes represent  the posterior-weighted EDF predictive distribution.} \label{JL-latent-vs-EDF}
\end{figure}

To visualize the discrete posterior and to construct continuous credible regions in the $(J,L)$ plane, we employ a kernel density estimation (KDE) procedure \cite{Silverman1986,Scott2015}. KDE does not introduce a physical assumption about the equation of state; it is used solely as a non-parametric visualization and interpolation tool.
The KDE-defined density $\hat p(J,L)$ is then used to construct highest-posterior-density (HPD) regions.
The 68\% (90\%) HPD region is defined as the smallest region in the $(J,L)$ plane that contains 68\% (90\%) of the total posterior probability,
\begin{equation}
\int_{\hat p(J,L)\ge p_*} \hat p(J,L)\,dJ\,dL
=
0.68\ (0.90),
\end{equation}
where $p_*$ is the corresponding density threshold.
By construction, all points inside a given HPD contour have higher posterior density than any point outside the contour.
To highlight the constraining power of the Sn neutron-skin systematics, we also compute the prior KDE obtained from the same EDF ensemble but with uniform weights.
A comparison between the prior and posterior densities makes explicit how the latent Sn constraint compresses the allowed region in $(J,L)$ space, particularly along the symmetry-energy slope parameter $L$.

\begin{figure}[t]
\includegraphics[width=8.8cm]{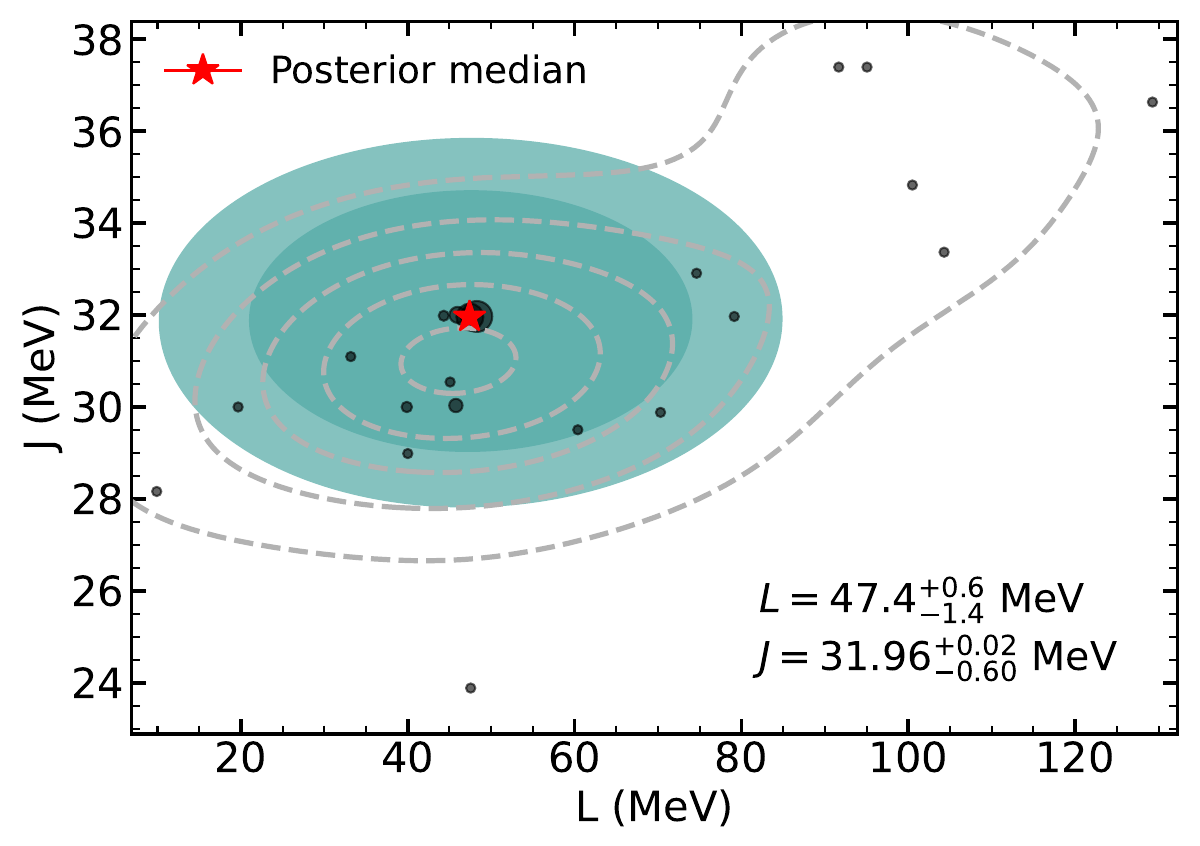}
\caption{Highest-posterior-density (HPD) constraints in the $(J,L)$ plane obtained from the latent Sn neutron-skin systematics.
Gray dashed contours show the unweighted EDF prior density in $(J,L)$ space.
Blue shaded regions denote the 90\% (light) and 68\% (dark) HPD posterior regions obtained after weighting each EDF by its likelihood against the latent Sn constraint.
Black points indicate individual EDFs, with marker size proportional to their posterior weight.
The red star marks the posterior median.
}
\label{JL-latent-vs-EDF-contour}
\end{figure}

Our results are shown in Figs. \ref{JL-latent-vs-EDF} and \ref{JL-latent-vs-EDF-contour}. The solid curve in Fig.  \ref{JL-latent-vs-EDF} and its shaded band show the inferred latent neutron-skin systematics along the Sn isotopic chain (same as Fig. \ref{fig:sn_chain}). The gray envelopes represent  the posterior-weighted EDF predictive distribution.
The close agreement between the latent band and the weighted EDF envelope indicates that a subset of EDFs reproduces the inferred skin systematics, enabling a calibrated mapping from the latent Sn trend to constraints on the associated nuclear-matter parameters  $(J,L)$. The envelope is tightest near stable Sn where the latent uncertainties are smallest and widens toward the chain ends where extrapolation and surface/isovector model differences become more important. The pronounced narrowing in $L$ reflects the strong sensitivity of neutron skins to the symmetry pressure, while the residual extent in $J$ arises primarily from correlations intrinsic to the EDF ensemble. The inferred median values and uncertainties are
\begin{align}
J &= 31.96~\text{MeV}\quad [68\%:\,31.36,\,31.98]~\text{MeV}, \nonumber\\
L &= 47.41~\text{MeV}\quad [68\%:\,46.03,\,48.02]~\text{MeV}.
\end{align}
These numbers are conditional on the EDF ensemble used; i.e., they reflect both the latent constraint and the model-class correlations built into that EDF set. We have verified that excluding PREX and CREX data from the analysis does not significantly alter the inferred symmetry-energy slope. This reflects the fact that the constraint on  $L$ is primarily driven by the systematic behavior of neutron skins across the Sn isotopic chain rather than individual measurements.

Fig. \ref{JL-latent-vs-EDF-contour}  shows the HPD constraints in the $(J,L)$ plane obtained from the latent Sn neutron-skin systematics. Gray dashed contours show the unweighted EDF prior density in $(J,L)$ space. Blue shaded regions denote the 90\% (light) and 68\% (dark) HPD posterior regions obtained after weighting each EDF by its likelihood against the latent Sn constraint. Black points indicate individual EDFs, with marker size proportional to their posterior weight. The red star marks the posterior median. 
The top-weight EDFs (by posterior weight) came out dominantly SLy5, SLy6, SLy4, SLy7 and SKM*. The reason is that our latent band is narrowest near stable Sn (around $A\approx 120$
($N \approx 70$). Since the EDF likelihood uses $\sigma(N)$ in the denominator, those nuclei dominate the fit $\chi^2 \approx \sum \Delta^2/\sigma^2(N)$.  Therefore,  the fit is ``anchored" there, while the neutron-rich end provides additional leverage but is somewhat down-weighted if  $\sigma(N)$ grows. 

\section{Conclusions and Outlook}

In this work we have developed and applied a hierarchical Bayesian framework to synthesize neutron-skin thickness constraints obtained from a wide range of experimental probes, theoretical correlations and astrophysical observations. By explicitly introducing method-dependent bias parameters and intrinsic nuisance widths, the analysis enables a statistically consistent combination of heterogeneous datasets without resorting to {\it ad hoc} error rescaling or selective data exclusion. This approach resolves long-standing ambiguities in global neutron-skin systematics while providing  a transparent assessment of both experimental and theoretical uncertainties.

A central result of this study is the inference of a smooth latent neutron-skin systematics as a function of isospin asymmetry and nuclear size, grounded in droplet-model and geometric considerations. The resulting posterior-predictive distributions along the Sn isotopic chain, shown in Fig.~\ref{fig:sn_chain}, display the expected monotonic growth of $\Delta r_{np}$ with neutron number, driven primarily by the increasing symmetry pressure in neutron-rich systems. The uncertainty bands exhibit a characteristic pattern: they are narrowest near stable Sn isotopes ($A \approx 120$), for which multiple probes provide overlapping constraints and widen toward both the proton-rich and neutron-rich extremes where extrapolation in isospin becomes unavoidable. This behavior reflects not only the density of available data but also the intrinsic structure of the latent model, in which uncertainties in the isovector slope parameters amplify at large asymmetry.

The hierarchical treatment provides a quantitative and data-driven ranking of neutron-skin probes. As summarized in Table~\ref{tab:method_systematics}, no method exhibits a statistically significant global bias relative to the chosen reference scale. Instead, apparent discrepancies among probes are explained by differing intrinsic scatters, which encode unmodeled theoretical and experimental systematics. Dipole-response observables (GDR, PDR, and AGDR) are particularly powerful in constraining the shape of isotopic trends, despite their reliance on EDF or RPA correlations. Antiprotonic atoms provide robust sensitivity to surface neutrons with moderate scatter, while parity-violating electron scattering, although theoretically clean, currently has limited leverage due to the small number of measured nuclei. Astrophysical constraints, when incorporated in a properly hierarchical and non-double-counted manner, act as weak but important regulators of the global trend rather than dominant drivers. These findings highlight the necessity of hierarchical modeling for avoiding overconfident or biased inferences.

Using the inferred latent Sn neutron-skin band as an empirical intermediary, we established a calibrated bridge to nuclear energy-density functionals. Weighting a broad ensemble of Skyrme EDFs by their agreement with the latent Sn systematics yields a posterior EDF envelope that closely tracks the inferred neutron-skin trend (Fig.~\ref{JL-latent-vs-EDF}), demonstrating internal consistency between nuclear structure data and a subset of EDFs. The resulting constraints in the $(J,L)$ plane, shown in Fig.~\ref{JL-latent-vs-EDF-contour}, reveal a pronounced compression of the symmetry-energy slope parameter $L$, while the constraint on the symmetry energy at saturation $J$ remains comparatively weaker. This outcome reflects the underlying physics: neutron skins are dominantly sensitive to the symmetry pressure at sub-saturation densities, which is governed by $L$, whereas sensitivity to $J$ arises mainly through correlations internal to specific EDF families. Importantly, the constraint on $L$ is driven by the isotopic dependence of $\Delta r_{np}$ across the Sn chain, rather than by any single nucleus, demonstrating the superior discriminatory power of isotopic trends over isolated benchmark measurements.

The methodology developed here provides a general blueprint for combining heterogeneous nuclear observables within a principled Bayesian framework. As experimental precision improves and astrophysical data continue to mature, such hierarchical approaches will be essential for transforming diverse measurements into robust constraints on the nuclear equation of state and for strengthening the quantitative bridge between finite nuclei and neutron-star matter.

\appendix
\section{Robustness of the Inferred Symmetry-Energy Constraints}

In this appendix we examine the robustness of the inferred neutron-skin systematics and the resulting constraints on the symmetry-energy parameters \(J\) and \(L\) under variations of model assumptions and dataset composition.

{\it Dependence on the neutron-skin parametrization.} The main analysis employs the leading-order parametrization
\begin{equation}
\Delta r_{np} = \beta_0 + \beta_1 I + \beta_2 A^{1/3} + \beta_3 I A^{1/3},
\end{equation}
which captures the dominant dependence on isospin asymmetry and nuclear size.

To assess the impact of higher-order contributions, we tested extensions including quadratic terms in \(I\) and additional surface-dependent corrections. We find that the inclusion of such terms does not significantly modify the inferred posterior distributions of \(J\) and \(L\). This indicates that the leading-order parametrization captures the dominant physics relevant for the present analysis, while residual effects are absorbed by the hierarchical nuisance parameters.

{\it Impact of dataset composition.} We evaluated the sensitivity of the results to the inclusion of specific datasets, in particular the parity-violating electron scattering measurements PREX and CREX.

Repeating the analysis with these data removed yields symmetry-energy constraints consistent with those obtained in the full dataset. This demonstrates that the inferred value of \(L\) is primarily driven by the systematic behavior of neutron skins across the Sn isotopic chain, rather than by individual measurements.

{\it Effect of model-dependent systematics.} The hierarchical model incorporates method-dependent bias parameters \(b_m\) and intrinsic nuisance widths \(\tau_m\), allowing for systematic offsets and underestimated uncertainties in different experimental probes.

To test the stability of the inference, we examined the posterior distributions under variations of the priors assigned to these nuisance parameters. The resulting changes in the inferred neutron-skin systematics and symmetry-energy parameters were found to be within the quoted uncertainties, indicating that the conclusions are not driven by specific prior choices.

{\it Dependence on the EDF ensemble.} The mapping from neutron-skin systematics to symmetry-energy parameters is performed using a representative set of non-relativistic energy density functionals. While the present analysis does not include relativistic mean-field models, we expect their inclusion to primarily broaden the posterior distributions rather than shift their central values.

This expectation is supported by the fact that the dominant sensitivity of neutron skins is to the symmetry-energy slope \(L\), which controls the pressure of neutron-rich matter near saturation density. Variations in model families therefore mainly affect the width of the inferred constraints rather than their qualitative behavior.

These tests demonstrate that the main conclusions of this work-namely, that neutron-skin systematics across the Sn isotopic chain provide a robust constraint on the symmetry-energy slope \(L\) are stable with respect to variations in parametrization, dataset composition, and modeling assumptions.

\bigskip

{\bf Acknowledgments}

This work was supported by U.S. Department of Energy Office of Nuclear Physics under Contract No. DE-SC0026074 with East Texas A\&M University, by the ExtreMe Matter Institute EMMI at the GSI Helmholtzzentrum f\"ur Schwerionenforschung, and by the European Union’s Horizon Europe Research and Innovation (EURO-LABS) programme under Grant Agreement No 101057511 .

\bibliographystyle{elsarticle-num}

\end{document}